\documentclass{raa}
\usepackage[numbers]{natbib}
\usepackage{graphicx,url,longtable,natbib} 

\begin{document}

\title{The kinematical and space structures
of IC 2391 open cluster and moving group with Gaia-DR2}
\volnopage{Vol.0 (200x) No.0, 000--000}
\author{E. S. Postnikova, \inst{1}
W. H. Elsanhoury, \inst{2,3}
Devesh P. Sariya, \inst{4}
N. V. Chupina, \inst{1}
S. V. Vereshchagin, \inst{1}
\and Ing-Guey Jiang \inst{4}
 }

\institute{Institute of Astronomy of Russian Academy of Sciences (INASAN),
48 Pyatnitskaya st., Moscow, Russia; \\
\and Astronomy Department, National Research Institute of Astronomy and Geophysics (NRIAG)
11421, Helwan, Cairo, Egypt (Affiliation ID: 60030681); \\
\and Physics Department, Faculty of Science, Northern Border University, Rafha Branch, Saudi Arabia; \\
\and Department of Physics and Institute of Astronomy,
National Tsing Hua University,
Hsin-Chu, Taiwan, {\it deveshpath@gmail.com};\\
}

\def\fhg{\hbox{$.\!\!^h$}}
\def\fsg{\hbox{$.\!\!^s$}}
\def\fdg{\hbox{$.\!\!^\circ$}}
\def\fmdeg{\hbox{$.\!\!^'$}}

\abstract
{The kinematical parameters, spatial shape and structure of the
open cluster IC 2391 and the associated stellar stream are studied here
using Gaia-DR2 (GDR2) astrometry data.
The apex positions are determined for the open cluster
IC 2391 (data taken from Cantat-Gaudin et al.) 
and for the kinematical stream's stars mentioned in Montes et al.
using both convergent point and AD-diagram methods.
The values of apex coordinates are:
$(A, D)_{CP}$=($6\fhg17\pm 0\fhg004, -6\fdg88\pm 0\fdg381$; for cluster)
\&
($6\fhg07\pm0\fhg007$, $-5 \fdg00\pm0\fdg447$; stream)
and
($A_0,D_0$)= ($6\fhg12\pm0\fhg004$, $-3\fdg4\pm0\fdg3$; cluster)
\&
($6\fhg21\pm0\fhg007$, $-11\fdg895\pm0\fdg290$; stream).
The results are in good agreement with the previously calculated values.
The positions of the stars in the disk and the spatial dispersion velocities are determined.
The paths of cluster and associated stream are traced in the disk by orbit calculation back
in time to their places of formation.
A possible genetic relationship between the cluster and the stream has been detected.
The approximation of the spatial and kinematical shape of the stream and the cluster is made.
According to this study, even though currently the cluster and the stream
seem to have spatial difference in their locations
but they appear to have formed in the same region of the Galactic disk.
}
\authorrunning{E. S. Postnikova et al.} 
\titlerunning{The structure of IC 2391} 
\maketitle

\section{Introduction}
Stellar moving groups are ensembles of gravitationally unbound stars moving with
almost identical space motions
(Montes et al. 2001, Chumak \& Rastorguev 2006).
IC 2391 is an interesting target which represents an open cluster
as well as a stellar moving group (a.k.a. stream, flow, or supercluster).
The young moving group IC~2391 was discovered by Eggen (1991).
In the Gaia era, with the availability of extremely precise data,
moving groups are seeing new limelight
to understand the Galactic disk in the solar neighborhood.
Also, young clusters like IC~2391 in the solar vicinity enable us to characterize
the abundance of our immediate portion of the Galactic disk
(D'Orazi \& Randich 2009).

IC~2391 (other names: MWSC~1529, Cl~VDBH~42, omi~Vel~Cluster, C~0838-528, Escorial~31)
is a nearby young open star cluster located in south of the Galactic plane
($\alpha_{2000}=08^h 40^m 32\fsg0$, $\delta_{2000}=-53^\circ 02^m 00^s$)
with Galactic coordinates $(l,b)=(270\fdg3622, -06 \fdg8387)$.

Due to its proximity, richness and
low reddening, the cluster is well-studied in a wide range of frequencies from
X-ray (Marino et al. 2005a),
optical (Pagano et al. 2009),
infrared (e.g. Siegler et al. 2007, Parker \& Tinney 2013)
to radio (Lim et al. 1996).
IC 2391 is known to harbour stars
which have just arrived on the zero-age main-sequence (Marino et al. 2005a).
The cluster consists of stars with spectral types ranging from B to M which
makes it interesting for studies of fast rotators
(Marino et al. 2005b). The cluster is expected to have lost a
significant amount of its population via evaporation caused by
dynamical processes (Boudreault \& Bailer-Jones 2009).

The cluster's heliocentric distance has been studied by various authors over the years.
Efremov et al. (1997) gave Hipparcos (ESA 1997)-based distance
modulus of IC~2391, $(m-M)=$ 5.84.
Using Hipparcos data for 11 stars, Robichon et al. (1999) obtained a distance of
$146^{+48}_{-45}$~pc.
Dodd (2004) evaluated the cluster's distance as $147\pm 5.5$~pc.
Barrado y  Navascu{\'e}s et al. (2004)
determined an age of 50 Myr for IC 2391
from the excess of lithium in the atmospheres of stars.
Platais et al. (2007) gave a slightly lower value (40 Myr) of its age using main-sequence fitting.
IC 2391 is known to have a low reddening value, $E(B-V)=0.01$ (Randich et al. 2001).

The proper motion of the cluster was given by Loktin (2003) as
($\mu_\alpha\cos\delta, \mu_\delta = -25.05\pm 0.34, 22.65\pm 0.28$~mas~yr$^{-1}$),
while Dodd (2004) gave mean proper motion components
$=-25.04\pm 1.53$ and $23.19\pm 1.23$~mas~yr$^{-1}$.
Both of these works used Tycho-2 catalogue (H{\o}g et al. 2000).
Dias et al. (2002) listed the value of the proper motion components as
$-24.97\pm 0.30$ and $22.70\pm 0.30$~mas~yr$^{-1}$.
The radial velocity of IC~2391 is $12.487\pm 3.533$~km~s$^{-1}$
as given by Conrad et al. (2014)
and about $14.49\pm 0.14$~km~s$^{-1}$ according to Dias  et al. (2002).

In the 3D velocity space, a kinematical stream (or, moving group)
of several dozens of stars
is associated with IC~2391 open cluster.
Eggen (1991) discovered the stream and used the term ``supercluster'' for it.
Montes et al. (2001) prepared a modern list of the member stars of IC~2391 stream.
The visible dimensions of the stream (in $V$) are $60$'$.00\times60$'$.00$.
The member stars of the stream occupy a broad area of the sky,
with its stars being scattered almost throughout the northern hemisphere.
Eggen (1991, 1995) suggested an age spread among the stream's stars.
Montes et al. (2001) provided an age estimate of 35 Myr for the stream.

With its unprecedented high accuracy,
the Gaia data allows us to address the question of the reliability
of the joint origin and the possible spatial-kinematic
connection of the star stream and cluster.
In this paper, we used two lists of stars:
cluster members of IC 2391 listed in Cantat-Gaudin et al. (2018)
and
another list of stars belonging to the stream from Montes et al. (2001).
Data for the cluster and the stream's stars can be found in Table~\ref{IC2391_tab1}
and Table~\ref{IC2391_tab2}, respectively.
We will use these parameters to determine apex positions (example, Vereshchagin et al. 2014;
Elsanhoury et al. 2016, 2018) for the cluster and the stream
using both convergent point method (CP) and AD-diagram method.
We also study the space motion, the velocity ellipsoids, the shape in the space and
birthplaces for both cluster and the stream.

The structure of this article is as follows: Section~\ref{OBS} explains the data used in this study.
Section~\ref{KINE} deals with the kinematical properties.
The next Section shows velocity ellipsoid parameters.
In Section~\ref{SHAPE}, we discuss the cluster's shape in space.
Section~\ref{BIRTH} gives the birthplaces of the IC~2391 cluster and the stream.
The conclusions of this work are presented in Section~\ref{CONC}.

\begin{table*}
\caption{Data for 39 stars of IC~2391 cluster.
The star list is taken from Cantat-Gaudin et al. (2018).
Apex positions and distance mentioned in the table are determined in this paper.}
\centering
\resizebox{\textwidth}{!}{
\begin{tabular}{cccccccccccccc}
\hline\hline\noalign{\smallskip}
GDR2&$\alpha_{J2015.5}$&$\delta_{J2015.5}$&$\pi$&$\sigma_\pi$&$\mu_\alpha$&$\sigma_{\mu_\alpha}$&$\mu_\delta$&$\sigma_{\mu_\delta}$&$V_r$&$\sigma_{V_r}$&$A$&$D$&d\\
&deg&deg&\multicolumn{2}{c}{mas}&\multicolumn{2}{c}{mas~yr$^{-1}$}&\multicolumn{2}{c}{mas~yr$^{-1}$}&\multicolumn{2}{c}{km~s$^{-1}$}&deg&deg&pc\\
\hline\noalign{\smallskip}
5317423293481147264&131.89258&-54.48348&6.55&0.02&-25.28&0.05&23.66&0.04&12.86&2.16&91.38&-1.04&151.65\\
5317884439832479872&130.75151&-53.90202&6.38&0.03&-23.29&0.06&22.85&0.06&16.48&4.82&94.29&-6.48&155.07\\
5317887321743547264&130.57572&-53.90217&6.64&0.04&-24.63&0.06&23.31&0.06&16.12&0.68&93.10&-6.34&148.84\\
5317906155187202176&130.34457&-53.63577&6.53&0.02&-24.91&0.04&23.20&0.05&16.51&1.32&92.58&-6.38&152.00\\
5318059532750974720&129.84384&-53.91815&6.48&0.05&-24.45&0.10&23.44&0.11&16.94&0.94&92.95&-6.86&151.61\\
5318069604459639552&129.10085&-54.01815&6.47&0.02&-23.69&0.04&23.38&0.05&14.11&2.21&91.02&-2.74&153.43\\
5318077507199948672&129.46469&-53.76262&6.68&0.02&-24.73&0.04&23.76&0.04&14.86&0.80&91.36&-4.04&148.63\\
5318093243960659456&130.20441&-53.62916&6.71&0.02&-24.04&0.04&23.80&0.04&13.55&0.61&91.99&-1.96&147.95\\
5318097916884923520&130.07602&-53.50790&6.60&0.02&-23.75&0.04&22.46&0.04&14.52&2.49&91.78&-4.32&150.34\\
5318150349846655488&128.93185&-53.35554&6.34&0.28&-24.26&0.53&23.51&0.50&17.61&7.77&92.55&-6.79&141.98\\
5318162238316886528&129.74460&-53.32008&6.60&0.02&-24.90&0.04&23.00&0.04&16.87&11.83&92.27&-7.11&150.20\\
5318170828251553792&129.31320&-53.33834&6.58&0.01&-24.02&0.03&23.07&0.03&9.66&12.47&87.13&4.84&151.29\\
5318176875565072768&129.22895&-53.14275&6.71&0.02&-23.96&0.05&23.78&0.04&16.20&1.52&93.07&-5.74&147.95\\
5318185667356683392&129.48414&-52.95143&6.66&0.03&-23.73&0.05&23.74&0.05&11.05&5.75&89.51&2.94&148.52\\
5318186221414047104&129.59528&-52.94657&6.66&0.02&-25.24&0.05&24.23&0.04&15.43&0.71&91.67&-3.76&148.91\\
5318229274167094656&131.56346&-53.75616&6.65&0.03&-24.66&0.05&22.92&0.04&16.00&0.83&93.73&-6.46&149.00\\
5318267653995965568&131.04233&-53.72592&6.68&0.03&-25.05&0.05&23.87&0.05&12.63&1.28&90.96&-0.33&148.43\\
5318296275658773888&131.44960&-53.43063&6.71&0.03&-26.09&0.05&23.48&0.05&15.83&8.73&92.44&-5.54&147.70\\
5318328676892604800&131.36203&-52.86715&6.65&0.02&-25.58&0.05&23.42&0.04&11.48&1.97&89.35&1.93&149.08\\
5318474941990522368&130.35768&-53.37808&6.65&0.03&-25.04&0.05&25.06&0.04&20.72&11.17&96.58&-10.53&149.08\\
5318504426950057728&130.49085&-52.87044&6.58&0.03&-25.73&0.05&22.70&0.05&17.12&2.11&92.07&-7.22&150.46\\
5318521671232021632&131.10882&-52.70888&6.58&0.03&-24.87&0.05&23.38&0.04&13.84&2.47&91.66&-1.62&150.78\\
5318532189619619328&130.98653&-52.68479&6.68&0.09&-22.95&0.16&22.84&0.14&14.84&3.70&94.33&-4.13&145.82\\
5318536999982933248&130.82442&-52.60302&6.52&0.03&-24.65&0.05&22.76&0.05&14.15&1.45&91.32&-2.43&151.65\\
5318541982138813824&129.99375&-53.05061&6.61&0.04&-25.30&0.07&24.69&0.07&17.24&0.83&93.48&-5.86&149.03\\
5318545521198976000&129.97087&-52.96569&6.56&0.02&-24.35&0.04&23.90&0.04&16.85&1.66&93.65&-5.87&151.25\\
5318546139674245888&129.92918&-52.96414&6.64&0.03&-25.44&0.05&22.48&0.05&16.78&2.79&91.52&-7.25&149.37\\
5318546822567826944&130.06753&-52.94134&6.54&0.03&-23.22&0.04&23.23&0.04&7.14&5.52&86.58&10.31&151.57\\
5318549678727766656&130.29139&-52.90283&6.58&0.03&-25.68&0.05&22.85&0.06&11.52&13.40&87.63&1.57&150.76\\
5318565656005527936&129.76157&-52.71059&6.50&0.02&-23.71&0.05&23.82&0.05&11.82&3.91&90.27&2.36&152.49\\
5318567958108066944&130.00665&-52.70338&6.64&0.03&-24.57&0.06&23.38&0.05&15.65&0.49&92.38&-4.63&149.21\\
5318647466543875584&131.91069&-52.26934&6.95&0.09&-25.41&0.16&24.42&0.18&16.68&2.97&95.37&-5.90&139.45\\
5321176205843037440&128.55421&-52.97214&6.70&0.02&-24.41&0.04&24.29&0.04&15.12&1.26&91.44&-3.43&148.14\\
5321188953307253760&129.35248&-52.90294&6.54&0.02&-23.94&0.05&22.51&0.05&15.50&4.71&91.55&-5.11&151.52\\
5321275295028390912&128.57546&-52.26594&6.44&0.02&-23.47&0.04&23.07&0.04&14.90&0.59&91.11&-2.80&153.80\\
5321280728169745536&128.75480&-52.23363&6.83&0.02&-24.52&0.04&24.36&0.04&17.26&2.55&93.34&-6.39&145.36\\
5321517672922567040&127.18998&-52.09067&6.42&0.02&-23.03&0.04&23.05&0.04&12.65&1.79&88.45&1.00&154.60\\
5321600445535344000&129.65022&-52.11067&6.63&0.03&-24.97&0.05&23.59&0.04&11.74&6.47&88.62&2.30&149.66\\
5321723109802924416&130.91001&-51.50794&6.54&0.05&-25.61&0.06&23.41&0.06&16.05&1.57&92.33&-3.85&150.41\\
\hline
\label{IC2391_tab1}
\end{tabular}
}
\end{table*}

\begin{table*}
\caption{Data listed for star stream in Montes et al. (2001)
cross-matched with other data sources to obtain astrometric parameters
and radial velocity.
Apex positions mentioned in the table are determined in this paper.
This table has 57 entries (see, Section~\ref{OBS}).
The values of parallaxes ($\pi$) and proper motions with their errors are taken from GDR2.
For the stars with no entries in GDR2 (with HIP numbers 11072 and 62686),
the values and errors of $\pi$ and proper motions are taken from van Leeuwen (2007).
Pairs of stars marked with an asterisk (*) near HIP number are considered likely double or binary stars.
The ``Ref.'' column contains the references for
$V_r$ and $\sigma_{V_r}$ according to:
a -- GDR2,
b -- Gontcharov (2006),
c -- Wilson(1963),
d -- Tokovinin(2013),
e -- Famaey et al.(2005) and
f -- White et al. (2007).}
\centering
\resizebox{\textwidth}{!}{
\begin{tabular}{cccccccccccccc}
\hline\hline\noalign{\smallskip}
HIP&GDR2&$\pi$&$\sigma_\pi$&$\mu_\alpha$&$\sigma_{\mu_\alpha}$&$\mu_\delta$&$\sigma_{\mu_\delta}$& $V_r$&$\sigma_{V_r}$&Ref.&$A$&$D$&d\\
&&mas&mas&\multicolumn{2}{c}{mas~yr$^{-1}$}&\multicolumn{2}{c}{mas~yr$^{-1}$}&\multicolumn{2}{c}{km~s$^{-1}$}&&deg&deg&pc\\
\hline\noalign{\smallskip}
4979&2538159890494125184&16.95&0.07&122.91&0.15&-40.36&0.07&5.6&2.90&         b&96.27&-17.73&58.44\\
 6869&2593154747696080640&19.39&0.06&112.68&0.11&-33.45&0.12&8.61&0.16&       a&91.00&-10.16&51.19\\
10175$^*$&77161217776670208&23.32&0.06&113.92&0.11&-72.09&0.10&25.09&0.21&    a&72.47&-12.91&42.64\\
10175$^*$&77161222072044288&23.26&0.05&111.01&0.09&-73.34&0.09&22.6&2.00&     c&74.31&-15.07&42.78\\
11072&--&45.53&0.82&196.61&0.81&-4.98&0.58&16.7&0.10&                         d&89.28&-15.85&21.07\\
12326$^*$&4741722617241539456&17.04&0.16&64.82&0.29&54.02&0.28&13.63&1.20&    a&81.44&-3.85&57.43\\
12326$^*$&4741722823399969536&17.28&0.03&74.68&0.05&48.74&0.05&16.45&0.24&    a&84.56&-9.90&57.66\\
12926&114575472461716864&39.60&0.06&237.61&0.09&-149.01&0.09&14.01&0.15&      a&95.98&-15.98&25.17\\
13081&114832620743735808&43.41&0.04&279.96&0.08&-119.15&0.06&10.21&0.21&      a&105.89&-11.52&22.99\\
14150&115311458058061440&47.16&0.06&233.12&0.12&-168.44&0.12&10.09&0.13&      a&100.29&-20.30&21.15\\
14954&3265335443260522112&44.37&0.20&193.25&0.32&-69.29&0.31&19.62&0.15&      a&94.88&-15.39&22.31\\
15058$^*$&3266941481859767680&13.79&0.13&80.89&0.18&-9.32&0.22&--&--&         a&--&--&71.00\\
15058$^*$&3266941486151475456&14.98&0.23&72.09&0.30&-15.07&0.39&28.3&0.20&    b&87.33&-6.90&64.42\\
22449&3288921720024442496&124.35&0.39&462.10&0.75&12.13&0.57&22.54&3.76&      a&110.75&6.40&7.98\\
23200&3231945508509506176&40.98&0.03&39.23&0.06&-95.05&0.04&18.08&0.63&       a&88.74&-28.80&24.36\\
25119&3234412606443085824&50.39&0.11&64.45&0.32&-180.09&0.24&36.30&0.24&      a&89.96&-22.15&19.74\\
29241&2912022740481002752&17.68&0.03&-28.65&0.04&45.06&0.05&26.98&0.19&       a&77.91&-1.46&56.31\\
33690&5479222240596469632&54.47&0.02&-162.07&0.05&264.64&0.04&22.04&0.19&     a&80.38&-13.75&18.34\\
40774&3089675232224086784&44.66&0.04&-164.23&0.07&-53.49&0.05&27.35&0.17&     a&92.44&-8.80&22.34\\
42253$^*$&666296182348971776&25.45&0.04&-110.14&0.07&-102.81&0.04&23.13&0.56& a&93.78&-13.46&39.15\\
42253$^*$&666296212412465024&25.54&0.04&-108.78&0.07&-102.91&0.05&19.30&0.29& a&90.61&-17.07&39.00\\
47193&1144716265940854016&3.72&0.29&-16.79&0.61&-17.61&0.50&-6.98&0.10&       e&98.90&-18.87&265.11\\
50371&5253347574048995584&4.34&0.39&-23.92&0.63&6.73&0.63&8.06&1.15&          a&85.82&-7.18&225.37\\
50660&749024502373761664&20.64&0.08&-152.52&0.09&-58.66&0.08&2.88&0.40&       a&80.59&-15.14&48.04\\
51931&3750851328223270400&31.36&0.07&-163.87&0.09&22.62&0.09&18.79&0.18&      a&106.70&-2.15&31.74\\
52468&5254185333222868352&3.40&0.25&-14.70&0.43&2.28&0.41&9.1&0.30&           b&90.33&-16.32&292.08\\
57198&4004885655800704896&3.94&0.07&-22.86&0.09&-7.49&0.08&5.14&0.28&         a&102.86&-12.05&253.84\\
59280&1536064958579187840&39.82&0.06&-314.20&0.05&-51.00&0.07&-2.61&0.12&     a&95.36&-9.57&25.03\\
60831&1541667932396172800&21.95&0.04&-182.13&0.04&-4.69&0.05&-2.21&0.18&      a&95.77&-3.31&45.37\\
60832&1541667932396172416&21.94&0.04&-180.39&0.05&0.44&0.06&-1.91&0.19&       a&94.92&-1.88&45.38\\
62686&--&26.13&3.38&-135.82&3.60&-27.22&5.02&-2.9&0.40&                       b&100.39&-12.89&27.23\\
62758&3958028490314315008&25.81&0.04&-141.07&0.06&-37.86&0.04&-4.24&0.23&     a&101.08&-17.35&38.59\\
66252&3630092241022731136&48.73&0.06&-286.58&0.11&-91.87&0.08&-23.16&0.16&    a&72.71&-8.44&20.46\\
67412&3658911226765436032&22.94&0.06&-136.86&0.09&-44.06&0.07&-13.44&0.25&    a&91.41&-15.52&43.33\\
68076&1671816367861537408&21.75&0.03&-138.54&0.06&54.51&0.05&-15.12&0.14&     a&89.90&-15.00&45.84\\
69713&1511727333122255744&34.30&0.18&-149.63&0.27&88.89&0.30&-18.4&2.70&      b&78.48&-12.79&28.80\\
74045&1701849283160198400&33.76&0.15&-127.83&0.31&165.69&0.30&-8.72&0.77&     f&83.02&-5.46&29.33\\
77152&1224551770875466496&20.67&0.05&-88.05&0.06&37.70&0.07&-20.5&0.35&       a&98.64&-4.14&48.14\\
77749&4403292246725472384&24.59&0.04&-115.09&0.07&10.81&0.07&-24.90&0.14&     a&99.94&5.00&40.52\\
84827&5811053234948685312&25.43&0.09&-47.08&0.13&-199.34&0.18&-3.30&0.20&     a&93.73&-14.47&39.00\\
85360&5976271757721062528&21.94&0.07&-26.13&0.12&-109.38&0.09&-16.03&0.27&    a&93.76&-22.43&45.28\\
89005&2260109892505203328&31.78&0.03&-26.17&0.06&194.87&0.06&-13.88&0.36&     a&99.42&-5.16&31.40\\
90004&4153637759337630720&24.03&0.06&-23.10&0.09&-68.30&0.09&-25.52&0.14&     a&104.78&-15.71&41.39\\
93096&6711686642605457408&13.53&0.04&24.30&0.06&-72.81&0.06&-12.82&0.40&      a&87.04&-17.56&73.41\\
99803$^*$&6468703708258513024&15.51&0.06&35.37&0.08&-98.19&0.07&-17.19&0.40&  a&106.35&-3.07&63.93\\
99803$^*$&6468703712555652096&15.45&0.05&37.07&0.08&-93.29&0.06&-19.67&0.44&  a&105.59&1.39&64.26\\
101262&1863898674120773760&37.13&0.03&141.86&0.05&16.81&0.06&-26.72&0.17&     a&90.15&-23.83&26.88\\
104225&2270771375724754816&31.00&0.03&108.41&0.05&66.26&0.05&-12.72&0.17&     a&86.75&-21.23&32.20\\
105232&1846882224145757056&25.45&0.06&133.59&0.11&9.56&0.11&-16.92&0.26&      a&82.42&-11.25&39.08\\
109110&2621051110038749440&27.45&0.05&176.76&0.09&-43.74&0.09&-11.31&0.63&    a&82.93&-11.15&36.27\\
109612&6508969718149502976&20.45&0.03&114.74&0.06&-65.22&0.06&-9.78&0.35&     a&97.15&-1.19&48.72\\
112909&1889703525209960960&66.42&0.07&523.78&0.10&-48.49&0.12&-2.78&0.55&     a&73.81&-6.72&15.02\\
113556&2663025241307454848&4.53&0.06&39.39&0.08&-6.67&0.08&-16.28&0.78&       a&95.69&-10.81&220.90\\
114236&6393744472971218048&17.80&0.04&103.61&0.05&-63.04&0.05&3.56&0.18&      a&103.06&-19.12&55.92\\
115288&2818000408811106688&15.61&0.18&73.65&0.37&20.25&0.22&-19&4.40&         b&121.99&0.46&62.40\\
116384&2646280705713202816&48.03&0.08&339.80&0.09&28.52&0.07&-10.44&0.52&     a&101.16&4.10&20.74\\
117410&2420563960807395072&35.47&0.90&234.22&1.45&23.16&1.30&-9.88&0.95&      a&103.04&9.14&26.58\\
\hline
\label{IC2391_tab2}
\end{tabular}
}
\end{table*}

\section{Data and sample of stars}
\label{OBS}

Using Gaia Data Release 2 (GDR2, Gaia Collaboration et al., 2016, 2018)
data, Cantat-Gaudin et al. (2018) provided a list of
224 probable members of IC 2391. Out of those, 39 stars have the availability of 
radial velocity information. 
This list of 39 stars is used in this paper 
for the analysis of IC 2391 cluster's kinematics.
The data used for cluster stars are listed in Table~\ref{IC2391_tab1}.

The distances to the stars are determined by the parallaxes $(\pi)$ from GDR2.
For each measurement of $\pi$,
the Monte Carlo method generated $N$ random variables
in the range of $(\pi\pm \sigma_\pi)$.
The resulting distribution is Gaussian with a maximum at a point with an argument value of $\pi$.
Artificially modeled parallax values are transferred to distances according to the formula
$1/\pi$, where $\pi$ is specified in arc seconds.
The distribution of the obtained distances is not normal.
By the method of least squares,
it is approximated by a curve representing the Maxwell distribution.
The argument values giving the maximum of this curve (maximum probability density)
are taken as the most likely distance values.

The list of stream stars was taken from Montes et al. (2001) and is shown in
Table~\ref{IC2391_tab2}.
The astrometric data (parallaxes and proper motions values) were
taken from GDR2 for most of the stars.
For two stars (HIP IDs 11072 and 62686),
the astrometric data were taken from van Leeuwen (2007).

Table~\ref{IC2391_tab1} and Table~\ref{IC2391_tab2} are compiled in a slightly different format:
in Table~\ref{IC2391_tab1} the first column gives GDR2 number,
while the first two columns in Table~\ref{IC2391_tab2} represent Hipparcos (HIP) and GDR2 numbers.
Then, the tables contain parallaxes ($\pi$) and their errors; proper motion components
($\mu_\alpha$, $\mu_\delta$) and their errors.
The next columns show radial velocities ($V_r$) and
their errors.
The references for the radial velocity data are mentioned in the ``Ref.'' column (Table~\ref{IC2391_tab2} only).
For Table~\ref{IC2391_tab1}, all the values of radial velocity are taken from Cantat-Gaudin et al. (2018).
The tables also present the values of apex (determined from AD-method) for individual stars determined in this work ($A, D$) and heliocentric distances ($d$).

\begin{figure}
\begin{center}
\includegraphics[width=10cm]{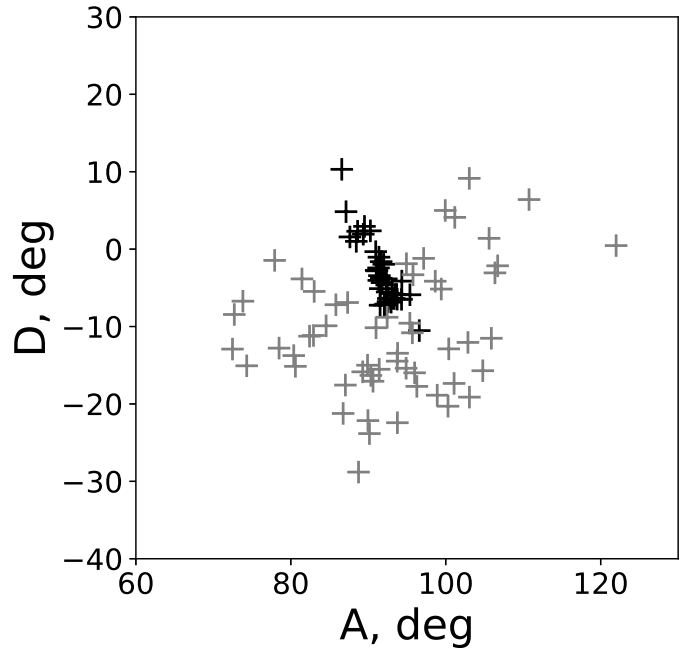}
\caption{The AD-diagram for IC 2391 open cluster and stream.
Black crosses represent data for 39 cluster stars from
Cantat-Gaudin et al. (2018) and the data for stellar stream from Montes et al. (2001)
is presented by gray crosses. The data for the stream contains 56 entries
considering that some stars could be double or binary stars.}
\label{ic2391_fig1}
\end{center}
\end{figure}

Montes et al. (2001) originally presented a list of 53 stars for the stream associated with IC 2391.
But, while cross-matching the data with Gaia DR2, it gave us two entries in GDR2 for some stars.
This is why Table~\ref{IC2391_tab2} has 57 entries of stars.
In the case of these stars,
for the same HIP IDs, there were different GDR2 IDs.
The fact that both the entries for these stars have approximately the same astrometric parameters
give support to the notion in the favor of them being likely double or binary stars.
The double star with two entries HIP~99803 was previously known.
While  the stars with Hipparcos IDs
HIP~10175, 12326, 15058, 42253 are new likely double or binary stars.
For the star with HIP ID 15058,
one of the entries had no available measurement for the radial velocity and its error.
This limited our calculations for apex to 56 entries.

Owing to its better spatial resolution and overall superior characteristics,
Gaia results are overtaking the Hipparcos ones in their
scientific significance.
This also means that in the regions where HIP could discover one star,
Gaia may discover multiple stars.
Although we are using the stars with multiple entries as likely double or binary stars
but we are not absolutely claiming about their status as double stars.
The close values of astrometric parameters made us include these stars into our list
rather than discarding that star entirely or choosing one of the two entries.

\begin{figure*}
\begin{center}
\includegraphics[width=14cm]{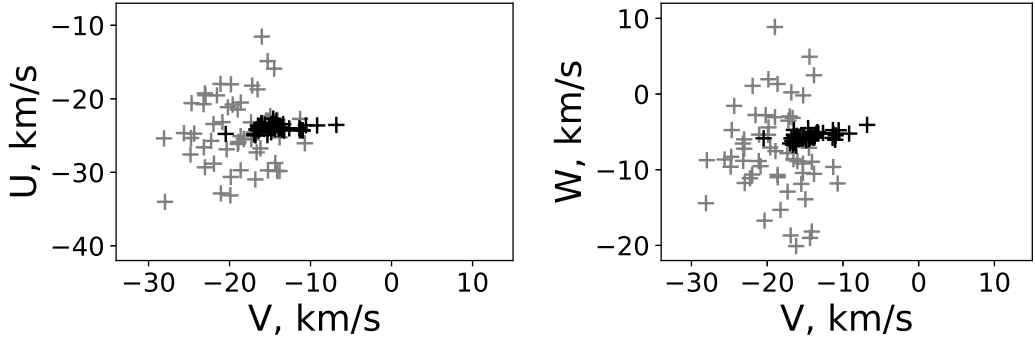}
\caption{Comparison of the spatial velocities of the stars of the cluster and the stream.
The velocities are given in km~s$^{-1}$. Cluster stars are shown in black color
and the stream stars are shown in gray.}
\label{ic2391_fig2}
\end{center}
\end{figure*}


\section{Kinematical properties}
\label{KINE}

\subsection{Apex (vertex position) of the cluster}

It is known that a moving cluster is a group of stars whose parallel motions
on the celestial sphere and its direction of proper motion will direct
towards a virtual point called convergent point or apex of this group.
This method has been used by our group to determine apex and other kinematical parameters for
open clusters M67, NGC 188 and Pleiades
(Vereshchagin et al. 2014, Elsanhoury et al. 2016, 2018).

Here, we calculate the apex for both cluster and stream represented by IC~2391
with the classical convergent point (CP) and the AD-diagram methods.

\subsubsection{The CP method}

It is a classical method which depends on the proper motion vectors components
(i.e. $\mu_\alpha\cos\delta$ and $\mu_\delta$ in mas~yr$^{-1}$).
Also, using the well-known formulae given by Smart (1968),
we can estimate the velocity components
$(V_x, V_y, V_z)$ along x, y, and z-axes
in the coordinate system centered at the Sun
for a group of $N$ cluster member stars with coordinates ($\alpha, \delta$),
at distance $r_i$ (pc), and with radial velocity $V_r$ (km~s$^{-1}$). i.e.


\begin{center}
$V_x= -4.74r_i\mu_\alpha\cos\delta\sin\alpha-4.74r_i\mu_\delta\sin\delta\cos\alpha
+V_r\cos\delta\cos\alpha$, \\
$V_y= +4.74r_i\mu_\alpha\cos\delta\cos\alpha-4.74r_i\mu_\delta\sin\delta\sin\alpha
+V_r\cos\delta\sin\alpha$, \\
$V_z=+4.74r_i\mu_\delta\cos\delta+V_r\sin\delta$.\\
\end{center}

From the above equations and letting
$\xi  = \frac{V_x}{V_z},$ $\eta = \frac{V_y}{V_z},$
we get
\[a_i \xi+b_i \eta  = c_i,\]
where the coefficients

\begin{center}
$a_i = \mu_\alpha^{(i)}\sin\delta_i\cos\alpha_i\cos\delta_i-\mu_\delta^{(i)}\sin\alpha_i$, \\
$b_i = \mu_\alpha^{(i)}\sin\delta_i\sin\alpha_i\cos\delta_i+\mu_\delta^{(i)}\cos\alpha_i$, \\
$c_i = \mu_\alpha^{(i)}\cos^2\delta_i$,
\end{center}

and the index $i$ varies from 1 to N which is the number of the cluster members. So
\begin{equation}\label{eq1}
\tan A_{CP} = \frac{\eta}{\xi},  \\
\end{equation}
\begin{equation}\label{eq2}
\tan D_{CP} = (\eta^2+\xi^2)^{-1/2}. 
\end{equation}

The coordinates $(A_{CP}, D_{CP})$ of the cluster apex are derived from
the Equations (\ref{eq1}) and (\ref{eq2}).
The results on the apex position of both the cluster and the stream are presented
in Table~\ref{IC2391_tab3}.

\subsubsection{The AD-diagram method}

The formulae to construct the AD-diagram can be seen in
Chupina et al. (2001, 2006).
The value of the apex coordinates for the IC 2391
cluster using data from Table~\ref{IC2391_tab1} is
$(A_0, D_0)= (6\fhg12\pm0\fhg004, -3\fdg4\pm0\fdg3)$.
For the stellar stream, using data from Table~\ref{IC2391_tab2}
the apex position is
$(A_0, D_0)=(6\fhg21\pm0\fhg007, -11\fdg895\pm0\fdg290)$.
Figure~\ref{ic2391_fig1} presents the constructed AD-diagrams for the cluster and the stream.
It is evident that the directions of movement in space
for the cluster and the stream almost coincide.

\subsection{Space velocity components ($U, V, W$)}

In order to compute the space velocity components $U$, $V$ and $W$,
we used an equatorial-Galactic transformation explained in Liu et al. (2011).
They determined the position of the Galactic plane
using recent catalogs like
Two-Micron All-Sky Survey (2MASS, Skrutskie et al. 2006) and
defined the optimal Galactic coordinate system
by adopting the ICRS position of the
compact radio source Sagittarius A$^*$ at the Galactic center (Liu et al. 2011).

The values of space velocity components along the three axes are given by:
\begin{center}
$U=-0.0518807421V_x\!-\!0.8722226427V_y\!-\!0.4863497200V_z$,\\
$V=+0.4846922369V_x\!-\!0.4477920852V_y\!+\!0.7513692061V_z$,\\
$W=-0.8731447899V_x\!-\!0.1967483417V_y\!+\!0.4459913295V_z$.
\end{center}

Figure~\ref{ic2391_fig2} shows the components ($U$, $V$, $W$) of the spatial
velocities of the cluster stars and the stream stars,
for which the input data were taken from
Tables~\ref{IC2391_tab1} and \ref{IC2391_tab2}.
As can be seen in Figure~\ref{ic2391_fig2},
the components of space velocities of all the stars do not show much different orientation.
Thus, it can be concluded that both the cluster and the stream
move in approximately the same direction.

\section{Velocity ellipsoid parameters}
\label{VELELP}

\subsection{Formulae for the Velocity Ellipsoid Parameters}

To compute the velocity ellipsoid and its parameters for IC~2391 open cluster
and the stream, we followed the computational algorithm mentioned in Elsanhoury et al. (2015).
A brief explanation of the algorithm is given here.

The coordinates of the $i^{th}$, star with respect to axes parallel to the original axes,
but shifted to the center of the distribution
i.e. towards average velocities ($\overline{U}$, $\overline{V}$ and $\overline{W}$),
will be $(U_i-\overline{U})$; $(V_i-\overline{V})$; $(W_i-\overline{W})$.
The average velocities $\overline{U}$, $\overline{V}$ and $\overline{W}$ are defined as:
\begin{equation}\label{mediumUVW}
\overline{U} = \frac{\sum\limits_{i=1}^{N} U_i}{N},\; \overline{V} = \frac{\sum\limits_{i=1}^{N} V_i}{N},\; \overline{W} = \frac{\sum\limits_{i=1}^{N} W_i}{N}.
\end{equation}
N being the total number of stars.

Let $\xi$ be an arbitrary axis, its zero point coincident with the center of
the distribution and let $l$, $m$ and $n$ be the direction cosines of the axis
with respect to the shifted ones, then the coordinates $Q_i$ of the point $i$,
with respect to the $\xi$-axis is given by:
\begin{equation}\label{Qi}
Q_i=l(U_i-\overline{U})+m(V_i-\overline{V})+n(W_i-\overline{W}).
\end{equation}
\begin{table*}
\caption{Kinematical parameters determined in this study for the IC~2391 cluster and the star stream.}
\centering
\begin{tabular}{ccc}
\hline\hline\noalign{\smallskip}
Parameters&Results&Reference\\
\hline\noalign{\smallskip}
$N$& 39 & Table~\ref{IC2391_tab1}\\
$N$& 56 & Table~\ref{IC2391_tab2}\\
$(A, D)_{CP}$& $6\fhg17\pm 0\fhg004$, $-6\fdg88\pm 0\fdg381$ & Table~\ref{IC2391_tab1}\\
& $6\fhg07\pm 0\fhg007$, $-5\fdg00\pm 0\fdg447$ & Table~\ref{IC2391_tab2}\\
$(A_0, D_0)$& $6\fhg12\pm 0\fhg004$, $-3\fdg4\pm 0\fdg3$ & Table~\ref{IC2391_tab1}\\
& $6\fhg21\pm 0\fhg007$, $-11\fdg895\pm 0\fdg290$ & Table~\ref{IC2391_tab2}\\
& $5\fhg83, -12\fdg44$ & Montes et al. (2001)\\
& $5\fhg82, -12\fdg44$ & Eggen (1991)\\
$\left(\overline{U},\overline{V},\overline{W}\right){\mbox{km s}^{-1}}$ & $-23.634, -14.449, -5.525$ & Table~\ref{IC2391_tab1}\\
& $-21.110, -7.213, -6.653$ & Table~\ref{IC2391_tab2}\\
& $-20.6, -15.7, -9.1$ & Montes et al. (2001)\\
$\left(x_c,y_c,z_c\right){\mbox{pc}}$ & $-57.84, 68.78, -119.66$ & Table~\ref{IC2391_tab1}\\
& $4.471, -0.552, 9.287$ & Table~\ref{IC2391_tab2}\\
Space velocity (km s$^{-1}$) & $28.25\pm0.19$ & Table~\ref{IC2391_tab1}\\
$ = \left(\overline{U}^2+\overline{V}^2+\overline{W}^2\right)^{1/2}$ & $23.28\pm 0.21$ & Table~\ref{IC2391_tab2}\\
& $27.453\pm 5.24$ & Montes et al. (2001)\\
& $30.00$ & Eggen (1991)\\
$\left(\lambda_1,\lambda_2,\lambda_3\right){\mbox{km s}^{-1}}$ & $779.54, 4.83, 0.19$ & Table~\ref{IC2391_tab1}\\
& $590.65, 44.06, 10.16$ & Table~\ref{IC2391_tab2}\\
$\left(\sigma_1,\sigma_2,\sigma_3\right){\mbox{km s}^{-1}}$ & $27.92, 2.20, 0.44$ & Table~\ref{IC2391_tab1}\\
& $24.30, 6.64, 3.19$ & Table~\ref{IC2391_tab2}\\
Dispersion velocity (km s$^{-1}$) & $28.11\pm0.18$ & Table~\ref{IC2391_tab1}\\
& $25.39\pm 0.20$ & Table~\ref{IC2391_tab2}\\
$\left(l_1,m_1,n_1\right){\mbox{deg}}$ & $0.83, 0.51, 0.20$ & Table~\ref{IC2391_tab1}\\
& $0.89, 0.34, 0.30$ & Table~\ref{IC2391_tab2}\\
$\left(l_2,m_2,n_2\right){\mbox{deg}}$ & $-0.53, 0.85, 0.01$ & Table~\ref{IC2391_tab1}\\
& $-0.36, 0.13, 0.92$ & Table~\ref{IC2391_tab2}\\
$\left(l_3,m_3,n_3\right){\mbox{deg}}$ & $0.16, 0.11, -0.98$ & Table~\ref{IC2391_tab1}\\
& $0.28, -0.93, 0.24$ & Table~\ref{IC2391_tab2}\\
$L_j,\, j=1,2,3$ & $-31\fdg655, -121\fdg757, 145\fdg685$ & Table~\ref{IC2391_tab1}\\
& $-20\fdg941, -159\fdg951, -106\fdg453$ & Table~\ref{IC2391_tab2}\\
$B_j,\, j=1,2,3$ &  $11\fdg286, 0\fdg511, -78\fdg702$ & Table~\ref{IC2391_tab1}\\
& $17\fdg490, 67\fdg342, 13\fdg946$ & Table~\ref{IC2391_tab2}\\
$S_{\odot}$,  km s$^{-1}$ &  $28.25\pm0.19$  & Table~\ref{IC2391_tab1}\\
& $23.28\pm 0.21$ & Table~\ref{IC2391_tab2}\\
$l_A $  & $-31\fdg439$ & Table~\ref{IC2391_tab1}\\
& $-18\fdg866$ & Table~\ref{IC2391_tab2}\\
$b_A$  & $11\fdg279$ & Table~\ref{IC2391_tab1}\\
& $16\fdg609$ & Table~\ref{IC2391_tab2}\\
\hline
\label{IC2391_tab3}
\end{tabular}
\end{table*}

If the measured scatter components are $Q_i$, a generalization of the mean square deviation can be defined as
\begin{equation}\label{mediumQ}
\sigma^2 = \frac{1}{N}\sum\limits_{i=1}^{N} Q^2_i.
\end{equation}
From equations (\ref{mediumUVW}), (\ref{Qi}) and (\ref{mediumQ}) we deduce
that
$\sigma^2=\underline{x}^TB\underline{x}$,
where $\underline{x}$ is the $(3\times1)$ direction cosines vector
and $B$ is $(3\times3)$ symmetric matrix with elements $\mu_{ij}$:
\begin{center}
$\mu_{11}=\frac{1}{N}\sum\limits_{i=1}^{N}U^2_i-(\overline{U})^2;\;\mu_{12}=\frac{1}{N}\sum\limits_{i=1}^{N}U_iV_i-\overline{U}\overline{V}$;\\
$\mu_{13}=\frac{1}{N}\sum\limits_{i=1}^{N}U_iW_i-\overline{U}\overline{W};\;\mu_{22}=\frac{1}{N}\sum\limits_{i=1}^{N}V^2_i-(\overline{V})^2$;\\
$\mu_{23}=\frac{1}{N}\sum\limits_{i=1}^{N}V_iW_i-\overline{V}\overline{W};\;\mu_{33}=\frac{1}{N}\sum\limits_{i=1}^{N}W^2_i-(\overline{W})^2$.
\end{center}
The necessary conditions for an extremum are now
\begin{equation}
\label{B_lam_I_0}
(B-\lambda I)\underline{x}=0. 
\end{equation}
These are three homogeneous equations in three unknowns,
which have a nontrivial solution if and only if
\begin{equation}
\label{det_lam}
D(\lambda)=\left|B-\lambda I\right|=0,
\end{equation}
where $\lambda$ is the eigenvalue, and $\underline{x}$ and $B$ are given as:

\begin{center}
$\underline{x}$=
$\left[
\begin{array}{c}
l\\
m\\
n
\end{array}
\right] $
and
$B$=
$\left[
\begin{array}{ccc}
\mu_{11}&\mu_{12}&\mu_{13}\\
\mu_{21}&\mu_{22}&\mu_{23}\\
\mu_{31}&\mu_{32}&\mu_{33}\\
\end{array}
\right] $
\end{center}

Equation (\ref{det_lam}) is the characteristic equation for the matrix $B$.
The required roots (i.e. eigenvalues) are
\begin{center}
$\lambda_1=\:\:2\,\rho^\frac{1}{3}\cos\frac{\phi}{3}-\frac{k_1}{3}$;\\
$\lambda_2=-\rho^\frac{1}{3}\left(\cos\frac{\phi}{3}+\sqrt{3}\sin\frac{\phi}{3}\right)-\frac{k_1}{3}$;\\
$\lambda_3=-\rho^\frac{1}{3}\left(\cos\frac{\phi}{3}-\sqrt{3}\sin\frac{\phi}{3}\right)-\frac{k_1}{3}$,
\end{center}
where
\begin{center}
$k_1\!=-\left(\mu_{11}+\mu_{22}+\mu_{33}\right)$;\\
$k_2\!=\mu_{11}\mu_{22}+\mu_{11}\mu_{33}+\mu_{22}\mu_{33}-\left(\mu^2_{12}+\mu^2_{13}+\mu^2_{23}\right)$;\\
$k_3\!=\mu^2_{12}\mu_{33}\!+\!\mu^2_{13}\mu_{22}\!+\!\mu^2_{23}\mu_{11}\!-\!\mu_{11}\mu_{22}\mu_{33}\!-\!2\mu_{12}\mu_{13}\mu_{23}$;\\
$q=\frac{1}{3}k_2-\frac{1}{9}k^2_1;\;\;\; r=\frac{1}{6}\left(k_1k_2-3k_3\right)-\frac{1}{27}k^3_1$;\\
$\rho=\sqrt{-q^3};\quad\quad x=\rho^2-r^2$;\\
$\phi=\tan^{-1}\left(\frac{\sqrt{x}}{r}\right)$.
\end{center}

Depending on the matrix that controls the eigenvalue problem
(Equation~\ref{B_lam_I_0}) for the velocity ellipsoid,
we establish analytical expressions of some parameters
for the correlations studies in terms of the matrix elements
$\mu_{ij}$ of the eigenvalue problem for the velocity ellipsoid
 (i.e. Velocity Ellipsoid Parameters, VEPs).

\begin{figure}
\begin{center}
\includegraphics[width=10cm]{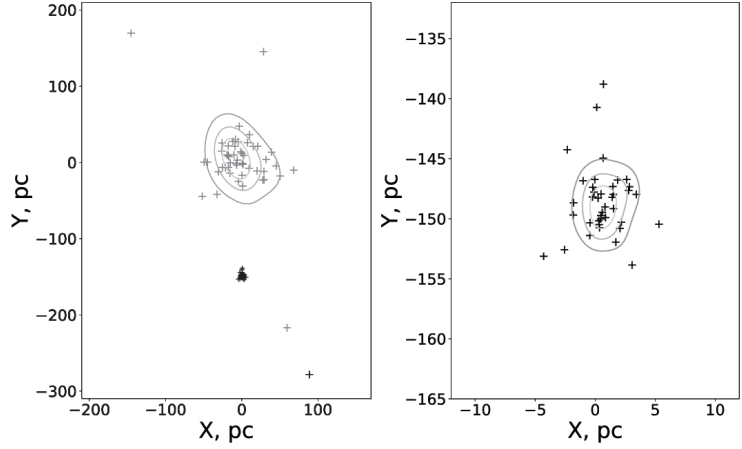}
\caption{
The Galactic XY plane distribution is shown here for the stream stars
(Table~\ref{IC2391_tab2}, gray crosses) -- left panel;
and for the cluster stars of IC~2391 (Table~\ref{IC2391_tab1}, black crosses) -- right panel.
In the left panel, the stars of the cluster are also visible as a small pile of black crosses at the lower side from
the center of the stream stars' distribution. Contours of equal stellar flux density are drawn in the both panels.
}
\label{ic2391_fig3}
\end{center}
\end{figure}

\subsubsection{The $\sigma_j, j\!=\!1,2,3$ parameters}

The $\sigma_j, j\!=\!1,2,3$ parameters are defined as $\sigma_j=\sqrt{\lambda_j}$.

\subsubsection{The $l_j$, $m_j$ and $n_j$ parameters}

The $l_j$, $m_j$ and $n_j$ are the direction cosines for eigenvalue problem.
We have the following expressions for $l_j$, $m_j$ and $n_j$ as
%
\begin{center}
$l_j =\frac{\mu_{22} \mu_{33} -\sigma_i^2\left(\mu_{22} +\mu_{33} -\sigma_i^2\right)-\mu_{23}^2}{D_j},\; j=1,2,3$;\\
$m_j =\frac{\mu_{23} \mu_{13} -\mu_{12} \mu_{33} +\sigma_{j}^{2} \mu_{12}}{D_j}, \; j=1,2,3$;\\
$n_j =\frac{\mu_{12} \mu_{23} -\mu_{13} \mu_{22} +\sigma_{j}^{2} \mu_{13}}{D_j}, \; j=1,2,3,$
\end{center}
where
\begin{center}
$D_j^2=\left(\mu_{22} \mu_{33} -\mu_{23}^{2} \right)^{2}+\left(\mu_{23} \mu_{13} -\mu_{12} \mu_{33} \right)^{2}$\\
$\!+\!\left(\mu_{12} \mu_{23} \!-\!\mu_{13} \mu_{22} \right)^{2}\!+\!2[\left(\mu_{22}\! +\!\mu_{33} \right)\left(\mu_{23}^{2}\! -\!\mu_{22} \mu_{33} \right)$\\
$\!+\!\mu_{12} \left(\mu_{23} \mu_{13} \!-\!\mu_{12} \mu_{33} \right)\!+\!
\mu_{13} \left(\mu_{12} \mu_{23}\! -\!\mu_{13} \mu_{22} \right)]\sigma_j^2$\\
$\!+\!\left(\mu_{33}^{2} +4\mu_{22} \mu_{33} +\mu_{22}^{2} \!-\!2\mu_{23}^{2} +\mu_{12}^{2} +\mu_{13}^{2} \right)\sigma_j^4$\\
$-2\left(\mu_{22} +\mu_{33} \right)\sigma_j^6 +\sigma_j^8$.
\end{center}
%

\begin{figure}
\begin{center}
\includegraphics[width=10cm]{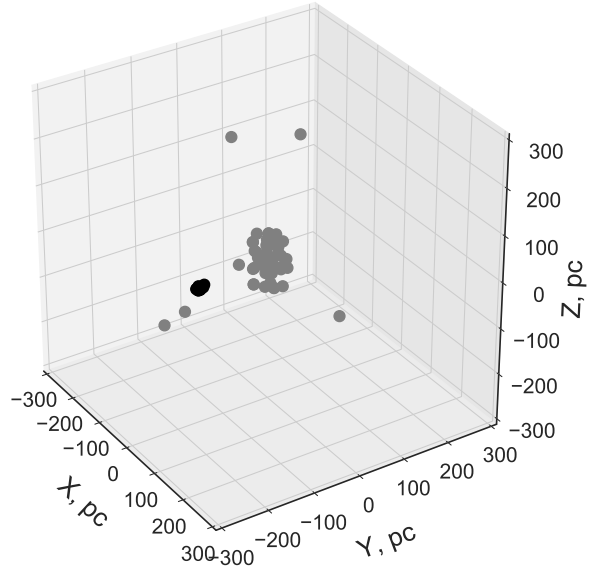}
\caption{The 3D picture of the cluster and stream stars in the rectangular Galactic coordinate system.
The black points are star cluster stars, while the gray points are the stream stars.}
\label{ic2391_fig4}
\end{center}
\end{figure}

Considering ($x_c, y_c, z_c$; pc) is the center of the cluster,
it can be estimated as the equatorial coordinates of the center of mass
for $N$ number of discrete objects as given below:
%
\begin{center}
$x_c=\frac{\sum\limits_{i=1}^{N}\left(r_i\cos\alpha_i\cos\delta_i\right)}{N}$, \\
$y_c=\frac{\sum\limits_{i=1}^{N}\left(r_i\sin\alpha_i\cos\delta_i\right)}{N}$, \\
$z_c=\frac{\sum\limits_{i=1}^{N}\left(r_i\sin\delta_i\right)}{N}$.
\end{center}
%

\subsubsection{The $L_j$ and $B_j$ Parameters}

Let $L_j$ and $B_j, j\! =\! 1, 2, 3$ be the Galactic longitude and the Galactic latitude of the directions which correspond to the extreme values of the dispersion, then
\[
L_j = \tan^{-1}\left(\frac{-m_j}{l_j}\right), \;\; B_j = \sin^{-1}(-n_j). 
\]

\subsubsection{The Solar elements}

The Solar motion can be defined as the absolute value of the Sun's velocity
relative to the group of stars under consideration, i.e.
\[
S_{\odot} = \left(\overline{U}^2+\overline{V}^2+\overline{W}^2\right)^{1/2}\!, \mbox{  km s}^{-1}.
\]
The Galactic longitude $l_A$ and Galactic latitude $b_A$ of the Solar apex are
\[
l_A = \tan^{-1}\left(\frac{-\overline{V}}{\overline{U}}\right), \;\; b_A = \sin^{-1}\left(\frac{-\overline{W}}{S_{\odot}}\right). 
\]

\subsection{VEPs calculated parameters}

We have a distribution of residual velocities of stars inside the cluster
and moving group.
In Table~\ref{IC2391_tab3} we present the results of our calculations.

\begin{figure*}
\begin{center}
\includegraphics[width=14cm]{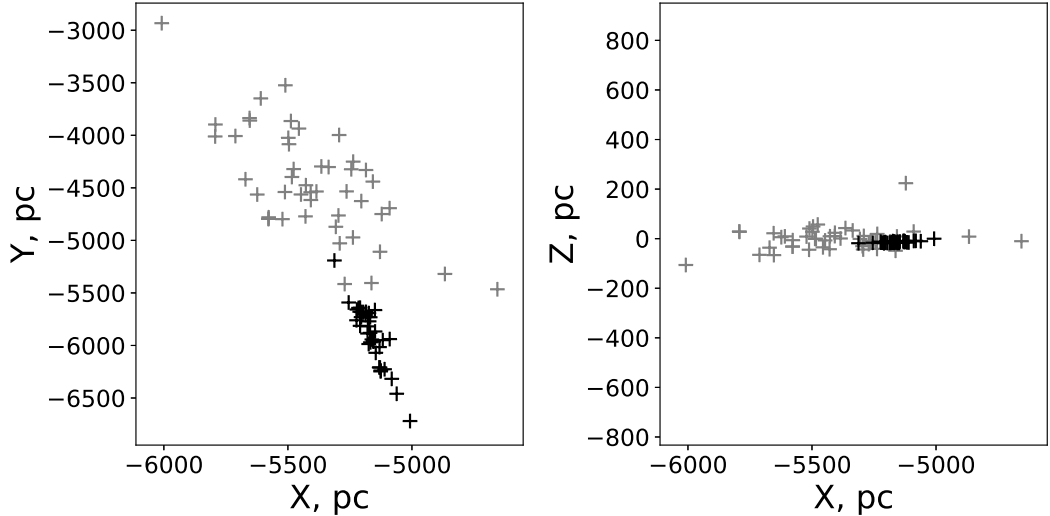}
\caption{Places of star formation for the cluster stars
(black crosses, data from Table~\ref{IC2391_tab1})
and for the stream stars (gray crosses, Table~\ref{IC2391_tab2}).
}
\label{ic2391_fig5}
\end{center}
\end{figure*}

\section{The cluster's and stream's shapes in space}
\label{SHAPE}

In Figure~\ref{ic2391_fig3}, the curves show contours of equal flux star density.
The kernel-density estimation was done considering Gaussian kernels (Scott, 2015)
using scipy python package (Jones et al. 2001) to make the isodensity plot.
At the periphery of the stream as well as the cluster too, stars are so few
that it is not possible to determine a significant value of density.
As can be seen in Figure~\ref{ic2391_fig3} (left panel),
the stars of the stream occupy a region stretched
roughly along (at a small angle not exceeding 40 degree) the axis OY.
The axis OY represents the direction of Galactic rotation.
The reason behind this orientation is the process leading to the decay of clusters
(from the associations containing multiple clusters)
with their gradual stretching (due to differential rotation)
along the direction of rotation of the Galactic disk.
This causes the gradual transformation of clusters into streams with their transformation
into ordered ring structures stretched around the Galactic center
(Perottoni et al. 2019, Wang et al. 2019).
Interestingly, a similar distribution of stellar flows/streams is observed in other galaxies
(Pearson et al. 2019).
The fact that the stream's stars are scattered over a wide area of the sky and there
are greater uncertainties in their selection, makes the distribution of stream stars
much more scattered than the cluster stars in Figure~\ref{ic2391_fig3}.
As for the IC~2391 cluster itself, represented by a handful of stars (from Table 1)
in Figure~\ref{ic2391_fig3} (right panel),
its spatial outlines shows the same pattern with a stretching along the OY direction.
Since the number of stars in the sample used here is small,
the structure is not defined in fine details.

Figure~\ref{ic2391_fig4} shows the 3D distribution of stars of the cluster and stream under consideration.
It is noticeable from Figure~\ref{ic2391_fig4} that the stars of the
cluster and the stream are located in separate regions in space.


\section{The birthplaces of the IC~2391 stream and open star cluster}
\label{BIRTH}

Figure~\ref{ic2391_fig5} shows the position of the stars of the cluster and the stream
approximately at the time of their formation,
determined by trial calculations of orbits back in time, up to 70 million years ago.
The galpy program was used to calculate the orbits (Bovy 2015).
Note that similar calculations carried out using another method are available
in Kharchenko et al (2009) and Chumak \& Rastorguev (2006).
Gravitational interactions among the stars of the stream were not taken into account.
For the cluster, the effect of irregular forces was not taken into account
(the dispersion of the peculiar velocities of stars, is a negligible value
compared with the spatial velocity).
In addition, the calculations did not take into account
the effect of spiral arms.
As suggested by Figure~\ref{ic2391_fig5},
we conclude that the birthplaces of the stars of the
cluster and the stream are in the same region of the disk.


\section{Conclusions}
\label{CONC}

In this work, we have determined various kinematical parameters of
the open cluster IC~2391 ($n=39$ stars) and the associated stream ($n=57$ stars).
A computational routine using the ``Mathematica'' software has been
developed to compute the kinematical structure.

We calculated the apex positions by two independent methods:
convergent point method and AD-diagram using the two data presented
in Table~\ref{IC2391_tab1} and Table~\ref{IC2391_tab2}.
We have determined the apex position with convergent point method
$(A, D)_{CP}$=($6\fhg17\pm 0\fhg004, -6\fdg88\pm 0\fdg381$; for cluster)
\& ($6\fhg07\pm0\fhg007$, $-5 \fdg00\pm0\fdg447$; for stream).
We have also obtained the apex values from AD-diagram
$(A_0, D_0)$ = ($6\fhg12\pm0\fhg004$, $-3\fdg4\pm0\fdg3$) and
($6\fhg21\pm0\fhg007$, $-11\fdg895\pm0\fdg290$)
for the cluster and the stream, respectively.
Results from both the convergent point and AD-diagram methods are almost similar.

For the cluster IC 2391 and the stream, we have determined the velocity ellipsoid parameters
(VEPs), including velocity dispersion ($\sigma_j$), 
direction cosines $l_j$, $m_j$ and $n_j$, 
Galactic longitude and latitude ($L_j$,$B_j$) with (j = 1, 2, 3), 
and the Solar elements  $S_{\odot}$, $l_A$ and $b_A$.
The parameters of the ellipsoids of residual velocities (VEPs) 
of the stars of the cluster and the stream are also determined.

It is evident from the two-dimensional Figure~\ref{ic2391_fig3} 
and the 3D structure of the cluster and the stream shown in Figure~\ref{ic2391_fig4} that 
the cluster and the stream are not spherically symmetric structures in the Galactic space. 
They are elongated in shape and directed along the direction of Galactic rotation.  
In general, all of this is consistent with the theoretical ideas 
regarding the tidal forces of the Galaxy acting on a stellar system, stretching the system towards the Galactic center 
and the differential rotation of the Galactic disk, turning the system in the direction of rotation. 
In addition, we obtained the evidence of the genetic connection between the 
star cluster IC 2391 and the stream back in time, leading to the time of their formation in the Galactic disk. 

\begin{acknowledgements}
We are thankful to the referee of this paper for useful and constructive comments.
Devesh P. Sariya and Ing-Guey Jiang are supported by the grant from
the Ministry of Science and Technology (MOST), Taiwan.
The grant numbers are
MOST 105-2811-M-007 -038,
MOST 105-2119-M-007 -029 -MY3, MOST 106-2112-M-007 -006 -MY3 and MOST 106-2811-M-007 -051.
This work has made use of data from the European Space Agency (ESA) mission {\it Gaia}
\footnote{https://www.cosmos.esa.int/gaia},
processed by the {\it Gaia} Data Processing and Analysis Consortium
(DPAC\footnote{https://www.cosmos.esa.int/web/gaia/dpac/consortium}).
Funding for the DPAC has been provided by national institutions,
in particular the institutions participating in the {\it Gaia} Multilateral Agreement.
This research has made use of the SIMBAD database, operated at
CDS\footnote{http://simbad.u-strasbg.fr/simbad/}, Strasbourg, France, Wenger et al. (2000).
\end{acknowledgements}

\end{document}